\documentclass[aps,12pt,preprintnumbers,nofootinbib]{revtex4}

\usepackage{amsmath,amssymb}
\usepackage{comment}
\usepackage{bm}

\usepackage{color}

\newcommand\x{\mathbf{x}}
\newcommand\y{\mathbf{y}}

\newcommand\q{\mathbf{q}}
\renewcommand\d{\partial}
\newcommand\+{\dagger}
\newcommand\Dlr{\raisebox{0.1em}{$\stackrel{\scriptstyle\leftrightarrow}D$}}
\newcommand{\bpm}{\begin{pmatrix}}
\newcommand{\epm}{\end{pmatrix}}

\preprint{EFI 14-42}
\begin{document}

\title{Particle-hole symmetry and electromagnetic response\\ of a
  half-filled Landau level} 
\author{Michael Levin and Dam Thanh Son}
\affiliation{James Franck Institute and
  Kadanoff Center for Theoretical Physics, University of Chicago, Illinois
  60637, USA}
\date{December 2016}

\begin{abstract}
We derive exact physical consequences of particle-hole symmetry of the
$\nu=1/2$ state of electrons in a strong magnetic field.  We show that
if the symmetry is not spontaneously broken, the Hall conductivity and
the susceptibility satisfy an exact relationship, valid at any wave
numbers and any frequencies much below the cyclotron frequency.  The
relationship holds for clean systems and also for systems with
statistically particle-hole symmetric disorder.  We work out the
constraints this relationship imposes on the theory of the Dirac
composite fermion.  We also argue that that the exact relationship is
violated in the Halperin-Lee-Read (HLR) field theory and present an explicit calculation
within a Galilean invariant mean-field approximation to the HLR theory
to illustrate the breakdown.
\end{abstract}


\maketitle

\section{Introduction}

The $\nu=1/2$ state of fermions in a strong magnetic
field~\cite{Halperin:1992mh} is one of the most important states in
quantum Hall physics.  It is at and near $\nu=1/2$ that sharp
predictions of the composite fermion (CF) theory were made and
successfully compared with
experiments~\cite{Willett:1993,Kang:1993,Goldman:1994zz}.  The
$\nu=1/2$ state is also the ``parent'' of the incompressible states in
the Jain sequences $\nu=n/(2n\pm1)$~\cite{Jain:1989tx} and of the
Moore-Read (MR) state~\cite{Moore:1991ks}.  

An important aspect of the $\nu =1/2$ quantum Hall system is that the
Hamiltonian is approximately symmetric under particle-hole (PH)
conjugation~\cite{Girvin:1984zz} in the spin-polarized lowest Landau
level.  In fact, this symmetry is exact for models with no Landau-level
mixing, complete spin polarization, and only two-body
interactions.
As an exact symmetry valid at all length scales, it must be present in
any low-energy, long-distance description.
At the same time, it is well known that the standard Halperin-Lee-Read (HLR) field theory~\cite{Halperin:1992mh} 
does not have any explicitly manifest symmetry that can be identified with the
particle-hole symmetry. 
To address this issue, an alternative field 
theory with explicit particle-hole symmetry has been proposed for
the $\nu=1/2$ state: the Dirac composite fermion
theory~\cite{Son:2015xqa}.  In this theory, the composite fermion is a
Dirac fermion, characterized by a Berry phase of $\pi$ around the Fermi
line.  Numerical simulations have confirmed this Berry
phase~\cite{Geraedts:2015pva}.

The Dirac composite fermion theory solves an old puzzle with the
theory of the composite fermion.  Particle-hole symmetry implies that,
when impurities are particle-hole symmetric (in the statistical
sense), the Hall conductivity $\sigma_{xy}$ is exactly
$\frac12(e^2/h)$.  In the HLR theory, this condition translates into a
Hall conductivity $-\frac12(e^2/h)$ of the composite
fermion~\cite{Kivelson:1997}.  This seems to contradict the fact that
the composite fermion feels zero average magnetic field.  If one takes
the Hall conductivity of the composite fermions to be zero, the
electron Hall conductivity is strictly less than $\frac12(e^2/h)$,
signaling the breakdown of particle-hole symmetry in the HLR theory.
Similarly, it was concluded in Ref.~\cite{Potter:2015cdn} that
thermoelectric transport in the HLR theory is also inconsistent with
particle-hole symmetry.
However, a more recent analysis shows that in a certain regime the CF
Hall conductivity may actually be $-\frac12(e^2/h)$~\cite{Wang:2016},
raising the question of whether the HLR theory is secretly particle-hole
symmetric.

In this paper we address the last question by deriving a consequence
of particle-hole symmetry for transport at nonzero wave number and
frequency.  We show that if the $\nu=1/2$ state
coincides with its own PH conjugate, then there exists an exact
relationship between two linear response functions, both regarded as
functions of wave number $q$ and frequency $\omega$: the Hall
conductivity and the susceptibility.  The relationship holds in the
presence of particle-hole symmetric disorder but remains nontrivial
in the absence of disorder.
Any low-energy effective theory of the half-filled Landau level must reproduce,
within its regime of validity, this exact relationship.
We then argue that the
relationship can be easily accommodated by the Dirac composite fermion
theory but is impossible to satisfy within the HLR theory.  This
rules out the possibility that the HLR theory has a hidden
particle-hole symmetry.\footnote{By the HLR theory we have in mind
  an effective field theory of a Fermi surface of composite fermions
  coupled to a Chern-Simons gauge field.  The composite fermions
  have nontrivial Landau parameters but have zero Berry phase.}

To write down the exact relationship, we first define the two
response functions.  Consider a small perturbation of the scalar
potential $A_0$.  Let $\delta \rho$ be the perturbation of the charge
density and $j^i$
be the current (more precisely, the ``$g=2$'' electromagnetic current
that remains finite in the lowest-Landau-level limit
$m\to0$~\cite{SimonSternHalperin:1996}).  The linear response of the
system to external $A_0$ is characterized by the susceptibility
$\chi(\omega,\q)$ and Hall conductivity $\sigma_H(\omega,\q)$
\begin{align}
  \delta \rho &= \chi (\omega,\q) A_0,\\
  j^i &= \left[\frac{\omega q^i}{q^2}\chi(\omega,\q)
       +i \sigma_H(\omega,\q) \epsilon^{ij} q_j \right] A_0. \label{j-subtracted}
\end{align}
Alternatively, the Hall conductivity can also be defined through the
density response to the perturbation of the magnetic field (the St\v reda
formula) when $A_0$ is left unperturbed,
\begin{equation}\label{Streda}
  \delta \rho = \sigma_H \delta B.
\end{equation}

Our result is that in the LLL limit and assuming that PH symmetry is
not spontaneously broken, the two response functions satisfy an exact
linear relationship,
\begin{equation}\label{exact-res}
  \sigma_H(\omega,q) + \frac1{4\pi}\tilde V(q)\chi(\omega,q) =
  \frac{1-e^{-q^2/2}}{2\pi q^2}\,.
\end{equation}
In this paper we set $B=1$, so the magnetic length $\ell_B=1$, and also
the CF Fermi momentum $p_F=1$.  The function $\tilde V(q)$ is fully
determined by the electron-electron interaction potential $V(r)$ or
its Fourier transform $V(q)$:
\begin{equation}\label{tildeV}
  \tilde V(q) = \frac2{q^2}(1-e^{-q^2/2}) V(q) - \frac2{q^2}
  \int\limits_0^\infty \! dp\, pe^{-p^2/2}[1-J_0(pq)]V(p).
\end{equation}
In particular, for Coulomb interaction $V(r)=e^2/(\epsilon_0r)$,
$\tilde V(q)$ can be computed exactly in closed form,
\begin{equation}\label{F-Coulomb}
  \tilde V(q) = \frac{4\pi e^2}{\epsilon_0}\biggl\{ \ \frac{1-e^{-q^2/2}}{q^3}
  - \frac1{q^2}
  \sqrt{\frac\pi 2} \biggl[ 1- e^{-q^2/4}I_0\biggl( \frac{q^2}4\biggr)
  \!\biggr] \biggr\}.
\end{equation}
Equation~(\ref{exact-res}) is valid for arbitrary wave numbers and for frequencies much smaller than the cyclotron frequency.
Equation~(\ref{exact-res}) also holds in the presence of PH symmetric impurities and at finite temperature, provided
that the thermal ensemble is PH symmetric.  This should be the case at
least at sufficiently high temperature, even if the potential $V(r)$
is such that the ground state breaks PH symmetry spontaneously. 

Equation (\ref{exact-res}) provides a nontrivial relationship between
two otherwise unrelated response functions.  One recalls that in a
Fermi liquid the response functions at finite $\omega$ have
singularities related to particle-hole pairs and other physical
excitations.  Since the right-hand side of Eq.~(\ref{exact-res}) does
not depend on $\omega$ at all, the singularities in $\sigma_H$ and
$\chi$ cancel exactly on the left-hand side of Eq.~(\ref{exact-res}).

If a state (denoted as ${\rm A}$) does not coincide with its
particle-hole conjugate (${\rm B}$), one can generalize
Eq.~(\ref{exact-res}) to relate the Hall conductivities of the two
states, $\sigma_H^{\rm A}$ and $\sigma_H^{\rm B}$ with the
susceptibility $\chi$ (which is the same in the two states),
\begin{equation}\label{exact-resAB}
  \frac{\sigma^{\rm A}_H(\omega,q) + \sigma^{\rm B}_H(\omega,q)}2
  + \frac1{4\pi}\tilde V(q)\chi(\omega,q) =
  \frac{1-e^{-q^2/2}}{2\pi q^2}\,.
\end{equation}


\section{Proof of the exact relation}

\subsection{Outline of the main argument}

We start from a microscopic theory describing spin-polarized electrons
of mass $m$, with gyromagnetic factor $g=2$, in an external magnetic
field, interacting through a two-body potential $V$:
\begin{multline}\label{H}
  H = \int\! d\x\, \Bigl[ \frac1{2m} |(\d_i - i A_i)\psi|^2
    - \Bigl( \frac B{2m}+A_0\Bigr) \psi^\dagger\psi \Bigr]
  \\ +
    \frac12 \int\! d\x\,d\y\, V(|\x-\y|) \psi^\+ (\x) \psi^\+(\y)
       \psi(\y) \psi(\x).
\end{multline}
The linear electromagnetic response of the system is given by $\delta
j^\mu(\omega,\q)=\Pi^{\mu\nu}(\omega,\q)A_\nu(\omega,\q)$, where
\begin{align}
  \Pi^{00}&=\chi,\\
  \Pi^{0i}&=\frac{\omega q^i}{q^2} \chi -i\epsilon^{ij}q_j\sigma_H,\\
  \Pi^{i0}&= \frac{\omega q^i}{q^2} \chi +i\epsilon^{ij}q_j\sigma_H.
\end{align}
The same function $\sigma_H$
governs the current response to the scalar potential and the density
response to perturbations of the magnetic field.  One can thus find
$\sigma_H$ by calculating the density of the ground state in
nonuniform magnetic fields (the St\v reda formula).

In constant $B$, PH conjugation flips the sign of
$A_0$~\cite{Levin:2007,SSLee:2007}.  We will show that in a nonuniform
magnetic field, the action of PH conjugation is more nontrivial: it
flips the sign of $A_0$ and simultaneously shifts it,
\begin{equation}\label{PHA0}
  \textrm{PH}: A_0 \to -A_0 + \delta_{\rm PH}A_0 .
\end{equation}
Here $\delta_{\rm PH}A_0$ is a functional of $B$ which vanishes when
$B$ is uniform.  For small perturbations of $B$, $B=B_0+\delta B$,
$\delta_{\rm PH}A_0$ is linear in $\delta B$.  For later convenience,
we parametrize the perturbation of the magnetic field through a
``K\"ahler potential'' $K$: $\delta B=\nabla^2 K$~\cite{Can:2014ota}.
Then $\delta_{\rm PH}A_0$ is linear in $K$, i.e.,
\begin{equation}\label{PHA0F}
  \delta_{\rm PH} A_0(\x) = -2 \int\!d\y\, F(\x-\y) K(\y) \equiv
  -2 F\ast K(\x),
\end{equation}
with some kernel $F$.

Another way to write Eq.~(\ref{PHA0}) is to define
\begin{equation}
  \tilde A_0 = A_0 - \frac12 \delta_{\rm PH}A_0[B].
\end{equation}
Then particle-hole conjugation simply flips the sign of $\tilde A_0$.
In particular, if $\tilde A_0=0$ or
\begin{equation}\label{deltaBA}
  A_0=\frac12\delta_{\rm PH}A_0[B] = - F\ast K,
\end{equation}
then the Hamiltonian is particle-hole symmetric.

Accepting the transformation law~(\ref{PHA0}), the argument leading to
Eq.~(\ref{exact-res}) goes as follows.  First, if the state under
consideration is PH symmetric in uniform magnetic field, then it will
remain PH symmetric under the small perturbation~(\ref{deltaBA}).
This means that the particle number density in this state is exactly
half of the density of the full Landau level in the (nonuniform)
magnetic field.  But the density is given by the linear response
formula
\begin{equation}
  \delta \rho=\Pi^{00}\delta A_0+\Pi^{0i}\delta A_i = \chi\delta A_0 +
  \sigma_H \delta B = \left(\sigma_H + \frac F{q^2}\chi\right)\delta
  B(q) .
\end{equation}
On the other hand, the density of the full Landau level is computable
to linear order in perturbations (see Ref.~\cite{Nguyen:2016itg} and
below),
\begin{equation}\label{n_nu=1}
  \delta \rho_{\nu=1}(q) = \frac{1-e^{-q^2/2}}{\pi q^2} \delta B(q).
  \end{equation}
This leads to Eq.~(\ref{exact-res}), with
\begin{equation}\label{VF}
  \tilde V(q) = \frac{4\pi}{q^2} F(q).
\end{equation}
Note that the coefficient on the right-hand side of Eq.~(\ref{n_nu=1})
is simply the Hall conductivity of the full Landau level,
\begin{equation}\label{sigmaH_nu=1}
  \sigma_H^{\nu=1}(q) = \frac{1-e^{-q^2/2}}{\pi q^2}\,.
\end{equation}

If one is dealing with a state which is not its own particle-hole
conjugate, repeating the above procedure and remembering that the
susceptibility $\chi$ is invariant under particle-hole symmetry, one
can derive Eq.~(\ref{exact-resAB}).

\subsection{Particle-hole conjugation in a nonuniform magnetic field}
\label{sec:PH-nonuniform}

To derive~(\ref{PHA0}), we can limit ourselves to perturbations which
are translationally invariant in one Cartesian coordinate, chosen to
be $y$.  In the Landau gauge $A_x =0$, $A_y = x + K'(x)$, the
normalized LLL orbitals (which are degenerate with zero
energy~\cite{Aharonov:1978gb}) have the form
$\psi_k(x,y)=L_y^{-1/2}\psi_k(x)e^{iky}$, where $L_y$ is the size of
the box along the $y$ direction, and to linear order in $K$
\begin{equation}\label{perturbed-psi}
  \psi_k(x) = \frac{1}{\pi^{1/4}} e^{-(x-k)^2/2} [
    1-K(x) + \bar K(k)].
\end{equation}
Here the function $\bar K$ is obtained by smearing $K$ by a Gaussian,
\begin{equation}\label{smearing}
  \bar K(k) = \frac1{\sqrt\pi}\int\!dx\, e^{-(x-k)^2} K(x),
\end{equation}
which implies that the Fourier transforms of $K$ and $\bar K$ are
related by $\bar K(q) = e^{-q^2/4} K(q)$.  (We use $k$, $l$,
etc., for momenta along the $y$ direction and $p$, $q$, etc., for
momenta along the $x$ direction).

The density of the $\nu=1$ state in inhomogeneous magnetic field can be computed exactly to linear order in perturbation,
\begin{equation}
  \rho(x)=\sum_k|\psi_k(x)|^2 = \int\!\frac{dk}{2\pi}\,\frac1{\sqrt\pi}
   e^{-(x-k)^2}[1-2K(x)+2\bar{K}(k)] = 
  \frac1{2\pi}[1-2K(x)+2\bar{\bar K}(x)],
\end{equation}
where $\bar{\bar K}$ is the function $K$ smeared [as in
  Eq.~(\ref{smearing})] twice.  In momentum space,
\begin{equation}
  \delta \rho(q) = -\frac{K(q)}\pi(1-e^{-q^2/2}),
\end{equation}
which coincides with Eq.~(\ref{n_nu=1}).

In the $m\to0$ limit, the Hamiltonian can be projected to the LLL.
Assuming that $A_0$ is also translationally invariant along the $y$
direction, the projected Hamiltonian is
\begin{equation}
  H = \sum_k U_k c^\+_k c_k + \frac12 \sum_{klmn} V_{klmn}c^\+_k c^\+_l
       c_m c_n \,,
\end{equation}
with
\begin{align}
  U_k &= - \int\!dx\, A_0(x) \psi_k^2(x) , \label{Uk}\\
V_{klmn}
  &= \int\!dx_1\,dx_2\,dy\, V(x_1-x_2,y) \psi_k(x_1)\psi_l(x_2)
  \psi_m(x_2)\psi_n(x_1)
  \frac1{L_y} e^{-i(k-n)y}\delta_{k+l,m+n}\,.
\end{align}
Note that $V_{klmn}$ is real and $V_{klmn}=V_{nmlk}$.

We will assume $A_0$ is of the same smallness as $K$; therefore we
can replace the wave functions in Eq.~(\ref{Uk}) by the unperturbed
wavefunctions at $K=0$.  We find
\begin{equation}\label{UkA}
  U_k = - \tilde A_0(k) =
  -\!\int\!\frac{dq}{2\pi}\,e^{iqk-q^2/4}A_0(q),
\end{equation}
where $A_0(q)$ is the Fourier transform of $A_0(x)$. 

We now perform PH conjugation of the Hamiltonian: $c_k\to
c_k^\+$.~\footnote{The PH conjugation used here is a \emph{unitary} transformation which
is a product of the antiunitary particle-hole conjugation of
Refs.~\cite{Levin:2007,SSLee:2007}, spatial reflection ($P$)
$y\to-y$, and time reversal $T$.  The product $PT$ transforms wave
functions as $\Psi(x_i,y_i)\to\Psi^*(x_i,-y_i)$ and seems to be a symmetry of
all quantum Hall states proposed so far, although, strictly speaking, there
seems to be no reason it could not be spontaneously broken. 
For $PT$ symmetric states,
invariance under the unitary PH conjugation implies invariance under
the antiunitary PH conjugation, and vice versa.}
After normal ordering, one finds that the two-body potential
remains unchanged, but the one-body potential is modified,
\begin{equation}\label{Uph}
   U_k \to - U_k + \sum_l ( - V_{kllk} + V_{klkl}).
\end{equation}
We will evaluate explicitly the sum by inserting
wave functions~(\ref{perturbed-psi}) into the definition of $V_{klmn}$.
But even without calculating, since we know that the result must be
linear in $K(q)$ and respect translational invariance along the $y$
direction, we can write it as
\begin{equation}\label{sumV}
  \sum_l ( - V_{kllk} + V_{klkl})
  = 2 \!\int\!\frac{dq}{2\pi}\,e^{iqk-q^2/4} F(q) K(q).
\end{equation}
From Eqs.~(\ref{UkA}), (\ref{Uph}), and (\ref{sumV}), we conclude
under PH conjugation
\begin{equation}
  A_0(q) \to -A_0(q) - 2F(q) K(q),
\end{equation}
and so $F(q)$ is the Fourier transform of the function $F(\x)$
introduced in Eq.~(\ref{PHA0F}).

We now split $F(q)=F^{\rm H}(q)+F^{\rm ex}(q)$, where
\begin{align}
  - \sum_l  V_{kllk}
  &= 2 \!\int\!\frac{dq}{2\pi}\,e^{iqk-q^2/4} F^{\rm H}(q) K(q),\\
  \sum_l V_{klkl}
  &= 2 \!\int\!\frac{dq}{2\pi}\,e^{iqk-q^2/4} F^{\rm ex}(q) K(q),
\end{align}
and compute $F^{\rm H}$ and $F^{\rm ex}$ separately.  The calculation
is straightforward but somewhat tedious; readers who are not interested
in the details can skip to Eqs.~(\ref{FH}) and (\ref{Fex-int}).
First, for the Hartree term $F^{\rm H}$,
\begin{align}
  - \sum_l V_{kllk} &= -\frac1{L_y}\sum_l\int\!dx_1\,dx_2\,dy\,
    V(x_1-x_2,y)  \psi^2_k(x_1) \psi_l^2(x_2) \\
  & = -\int\!dx_1\,dx_2\,\frac{dl}{2\pi}\,
    V_1(x_1-x_2)\psi_k^2(x_1)\psi_l^2(x_2),
\end{align} 
where $V_1(x)=\int\!dy\,V(x,y)$.  Corrections of order $O(K)$ appear
in both $\psi_k(x_1)$ and $\psi_l(x_2)$, but it is easy to see that the
only nontrivial $O(K)$ contribution comes from $\psi_l(x_2)$.  We thus
have
\begin{equation}
   - \sum_l V_{kllk}=
  - \int\!dx_1\,dx_2\,\frac{dl}{2\pi}\,V_1(x_1-x_2)
     \frac1\pi e^{-(x_1-k)^2-(x_2-l)^2} 2[-K(x_2) + \bar K(l)].
\end{equation}
Integration over $l$ yields
\begin{equation}
  \frac1{\pi} \int\!dx_1\,dx_2\, V_1(x_1-x_2) 
  \frac1{\sqrt\pi} e^{-(x_1-k)^2}
  [K(x_2)-\bar{\bar K}(x_2)].
\end{equation}
Rewriting in momentum-space representation, one finds
\begin{equation}
  - \sum_l V_{kllk} 
= \frac1{\pi} \int\!\frac{dq}{2\pi}\,
   e^{-q^2/4} e^{iqk} V(q) 
(1-e^{-q^2/2}) K(q),
\end{equation}
which means
\begin{equation}\label{FH}
  F^{\rm H}(q)= \frac{1{-}e^{-q^2/2}}{2\pi} V(q).
\end{equation}

Now let us turn to the exchange contribution,
\begin{equation}
   \sum_l V_{klkl} = 
  \int\!dx_1\,dx_2\,dy\, \frac{dl}{2\pi}\, \psi_k(x_1)\psi_l(x_2)\psi_k(x_2)
     \psi_l(x_1) e^{-i(k-l)y} V(x_1-x_2, y).
\end{equation}
Expanding the wavefunctions to linear powers in $K$ using
Eq.~(\ref{perturbed-psi}), we get
\begin{multline}
  \sum_l V_{klkl} = \frac2\pi\! \int\!dx_1\,dx_2\,dy\,\frac{dl}{2\pi}
  e^{-\frac12\left[(x_1-k)^2+(x_2-l)^2+(x_2-k)^2+(x_1-l)^2\right]
     -i(k-l)y}
  \\
  \times
  V(x_1-x_2,y) [-K(x_1)-K(x_2)+\bar K(k)+\bar K(l)].
\end{multline}
To evaluate this integral, we represent $K$, $\bar K$, and the
potential $V$ in Fourier components,
\begin{equation}
  K(x) = \int\!\frac{dq}{2\pi}\, e^{iqx} K(q), \quad
  \bar K(k) = \int\!\frac{dq}{2\pi}\, e^{iqk-q^2/4} K(q),
\end{equation}
\begin{equation}\label{V-Gauss}
  V(x,y) = \int\!\frac{dp_x\, dp_y}{(2\pi)^2}\, e^{i(p_xx+p_yy)}
    \,V(p), \qquad p \equiv \sqrt{p_x^2+p_y^2}\,.
\end{equation}
After integrating over $y$ and $l$, we obtain
\begin{equation}
  \frac2\pi\int\!\frac{dq}{2\pi}\,\frac{dp_x\,dp_y}{(2\pi)^2}\, V(p) K(q)\!
  \int\!dx_1\,dx_2\,
  e^{S_0} \left(-e^{iqx_1}-e^{iqx_2}+e^{iqk-q^2/4}
     +e^{iq(k-p_y)-q^2/4} \right),
\end{equation}
where
\begin{equation}
  S_0 = -\frac12\left[ (x_1-k)^2+(x_2-k+p_y)^2+(x_2-k)^2+(x_1-k+p_y)^2\right]
        + ip_x(x_1-x_2).
\end{equation}
The integral over $x_1$, $x_2$ is a Gaussian integral which can be
evaluated exactly.  We get
\begin{equation}
  2\! \int\!\frac{dq}{2\pi}\,\frac{dp_x\,dp_y}{(2\pi)^2}\, V(p) K(q)
  e^{-p^2/2}e^{ikq-q^2/4}(-e^{-(p_x+ip_y)q/2}-e^{(p_x-ip_y)q/2}+1
  +e^{-ip_yq}) .
\end{equation}
Going to polar coordinates in the $(p_x,p_y)$ plane and integrate over the
angle, we find
\begin{equation}
  \sum_lV_{klkl} = - 2\!\int\!\frac{dq}{2\pi}\!
  \int\limits_0^\infty\!\frac{dp}{2\pi}\,p 
  V(p) K(q) e^{-p^2}e^{iqk-q^2/4} [1-J_0(pq)],
\end{equation}
which implies
\begin{equation}\label{Fex-int}
  F^{\rm ex}(q) =  
   - \int\limits_0^\infty\!\frac{dp}{2\pi}\, pe^{-p^2/2}[1-J_0(pq)] V(p).
\end{equation}
Summing up $F^{\rm H}$ and $F^{\rm ex}$, and using Eq.~(\ref{VF}), we
find Eq.~(\ref{tildeV}).  The exact result (\ref{exact-res}) is now
proven.

Later, we will need the Taylor expansion of $F^{\rm ex}(q)$ over
$q^2$,
\begin{subequations}\label{Fex-series}
\begin{align}
  F^{\rm ex}(q) & = \sum_{n=1}^\infty \frac{(-1)^n}{n!} 
  \left( \frac{q^2}2\right)^n V_n\,, \label{Fex-series1}\\
  V_n &= \frac1{2^n n!}\int\limits_0^\infty\!\frac{dp}{2\pi}\,p^{2n+1}e^{-p^2/2}V(p)
  =
  \int\limits_0^\infty\!dr\, r L_n\left(\frac{r^2}2\right)e^{-r^2/2}
       V(r) .\label{Fex-series2}
\end{align}
\end{subequations}
In contrast to $\tilde V^{\rm H}(q)=4\pi q^{-2}F^{\rm H}(q)$,
which inherits the singularity at small $q$ of the potential (for
example, of the Coulomb potential), $\tilde V^{\rm ex}(q)=4\pi
q^{-2}F^{\rm ex}(q)$ is regular at $q=0$ for reasonably behaving
potentials.

For the Coulomb potential $V(r)=e^2/(\epsilon_0r)$, using the formulas
above, the function $F(q)$ can be evaluated exactly to be the one
given by Eqs.~(\ref{F-Coulomb}) and (\ref{VF}).

\subsection{Discussion}

Our formula~(\ref{exact-res}) does not determine $\sigma_H$ and $\chi$
separately and only fixes a linear combination of the two.  However,
even this limited statement can give very interesting results, which,
in principle, should be verifiable by numerical simulation of the
half-filled Landau level.

First, we notice that for a purely repulsive potential [$V(r)>0$ for
all $r$], $F(q)$ vanishes at some value of $q$.  In fact, at $q\to 0$,
one can show that
\begin{equation}
  F(0) = \frac{q^2}2 \int\!dr\, rV(r) \left[1-e^{-r^2/2}\left(1-
  \frac{r^2}2\right)\right]
\end{equation}
and hence is positive, while for $q\to\infty$ the exchange part
$F^{\rm ex}$ dominates, and according to Eq.~(\ref{Fex-int}), it
approaches a negative constant, which can be shown to be $F(\infty) =
- \int_0^\infty\!dr\,e^{-r^2/2}V(r)$.  At $q$ where $F(q)=0$, the Hall
conductivity is exactly determined by Eq.~(\ref{exact-res}).  For
example, for the Coulomb interaction $F(q)$ vanishes at $q=q_0\approx
1.4197\ell_B^{-1}$.  The Hall conductivity $\sigma_H$ at this
wave number can be predicted to be 0.3150 $e^2/h$ and is independent
of frequency.

Another interesting value of $q$ is $q=2$, corresponding to the $2k_F$
singularity in the response functions.  Our result implies that
although each function $\chi$ and $\sigma_H$ may show Friedel-type
singular behavior at this wave number, the linear combination
\begin{equation}
  \sigma_H + \frac14 F(2)\chi, \qquad F(2) \approx - 0.2372 \frac{e^2}{\epsilon_0}
\end{equation}
should be free of all singularities, for all $\omega$.  

Finally, in the limit $q\to0$, $\omega\to0$, $q/\omega\to0$, the
density-density correlation function is expected to behave as
$\chi(q)\sim q^4$ (the $q^2$ term in $\chi$ is fixed by Kohn's
theorem and vanishes in the LLL limit $m\to0$).  For any potential
which is less singular than $1/q^2$ at small $q$, the $q^2$ correction
to the ac Hall conductivity is then completely fixed,
\begin{equation}\label{sH-smallq}
  \sigma_H(\omega,q) = \frac1{4\pi}\left( 1- \frac{q^2}4 \right) + o(q^2),
  \qquad v_Fq \ll \omega 
\end{equation}
To order $q^2$, this is exactly one half of $\sigma_H$ for a
filled Landau level.

\section{Constraints on the Dirac composite fermion theory}

We now show that the exact relationship can be accommodated by the
Dirac composite fermion theory.  To illustrate how such a theory can
be constructed, we start with the simplest model Lagrangian and then
improve it.

Let us start from the action providing the dual description of Dirac
fermions~\cite{Metlitski:2015eka,Wang:2015qmt,Mross:2015idy,Karch:2016sxi,Seiberg:2016gmd}
\begin{equation}
  S = i\bar\psi\gamma^\mu(\d_\mu-ia_\mu)\psi - \frac1{4\pi} Ada,
\end{equation}
where $Ada\equiv\epsilon^{\mu\nu\lambda}A_\mu\d_\nu a_\lambda$ and, for
convenience, here and below integration over space-time is implied in
the action.  As we are interested only in local response, we ignore
issues related to the parity anomaly and the fractional coefficients
of the CS terms.  This action has a $CP$ symmetry, under which $x\to
x$, $y\to-y$ and
\begin{equation}\label{CT}
  A_0\to -A_0,  \quad A_x \to -A_x,\quad A_y \to A_y,
  \quad a_0\to a_0, \quad a_x\to a_x, \quad a_y \to -a_y .
\end{equation}
Under $CP$ the Hall conductivity flips sign; therefore any
$CP$ symmetric state must have zero Hall conductivity.

Consider now the half-filled Landau level with a long-range
electron-electron interaction.  At minimum, one has to add two more
terms to the action
\begin{equation}
  S = S_{\rm CF}(\psi, a) - \frac1{4\pi} Ada + \frac1{8\pi}AdA -
  \frac12  V  \delta\rho^2 ,
\end{equation}
with
\begin{equation}
  \delta\rho = \frac1{4\pi}(\delta B-b),
\end{equation}
being the density perturbation, and we use the shorthand notation
$V\delta\rho^2\equiv\int\!d\x\,d\y\,V(\x-\y)\delta\rho(\x)\delta\rho(\y)$.
The two terms added 
are the
$AdA$ term, which shifts the particle-hole symmetric value of
$\sigma_H$ from 0 to $\frac12(e^2/h)$, and the term containing $V$,
which describes the long-range density-density interaction.  The
additional terms violate the $CP$ symmetry~(\ref{CT}).  However, if
one separates out the Chern-Simons term $AdA$, which depends
exclusively on the background field,
\begin{equation}
  S = S_{\rm ph} (\psi,\tilde A,a) + \frac1{8\pi}AdA ,
\end{equation}
the remaining part $S_{\rm ph}$ can be put into the form
\begin{equation}
  S_{\rm ph} = S(\psi,a) - \frac1{4\pi}\tilde A da - \frac 12 \frac1{(4\pi)^2}
  ( V \delta B^2 +  V b^2),
\end{equation}
where
\begin{equation}
  \tilde A_0 = A_0 - \frac{V\ast\delta B}{4\pi}\,,\qquad \tilde A_i = A_i .
\end{equation}
Now one can see that $S_{\rm ph}$ has a modified $CP$ symmetry, under which
\begin{equation}\label{mod_CT}
  \tilde A_0\to -\tilde A_0,  \quad A_x \to -A_x,\quad A_y \to A_y,
  \quad a_0\to a_0, \quad a_x\to a_x, \quad a_y \to -a_y .
\end{equation}

To derive the consequences of this symmetry for transport, imagine
that we integrate out the dynamical $\psi$ and $a$.  We now have an
effective action for $A_\mu$ which is constrained by the
symmetry~(\ref{mod_CT}).  To the quadratic order
\begin{equation}
  S_{\rm eff}[A] = \frac\chi 2 \tilde A_0^2 - \frac{\chi_M} \delta B^2
     + \frac1{8\pi} AdA ,
\end{equation}
where $\chi$ is the susceptibility and $\chi_M$ is a coefficient
related to the magnetic susceptibility.  From this effective action it
is straightforward to derive
\begin{equation}
  \sigma_H + \frac V{4\pi} \chi = \frac1{4\pi} \,.
\end{equation}

This is similar to but not yet the exact
relationship~(\ref{exact-res}): instead of $\tilde V(q)$ we have only
the leading part $V(q)$, and on the right-hand side is just a constant
instead of the full function of $q$.  However, it is easy to modify
the action to reproduce correctly the exact relationship.  Let
$S_{\nu=1}[A]$ be the action describing the full lowest Landau level.
This is a complicated functional of $A$, but to the quadratic level it
is completely determined by the Hall conductivity~(\ref{sigmaH_nu=1}),
\begin{equation}
  S_{\nu=1}[A] = \frac1{4\pi} \epsilon^{\mu\nu\lambda}A_\mu\d_\nu A_\nu
  - \frac 1{8\pi} (\bm{\nabla}\cdot\mathbf{E})
    \frac{1-e^{\nabla^2/2}+\nabla^2/2}{\nabla^4/8} B .
\end{equation}

With a bit of hindsight, consider the following action:
\begin{equation}\label{S-improved}
  S = S_0(\psi,a, A) + \frac12 S_{\nu=1}[A]
  - \frac V2 \left(\frac{\delta S_0}{\delta A_0} + \frac12
  \frac{\delta S_{\nu=1}}{\delta A_0} -\rho_0 \right)^2
  + \sum_{n=0}^\infty C_n \frac{\delta S_0}{\delta A_0} \nabla^{2n} \delta B,
\end{equation}
where $S_0[\psi,a,A]$ is a local Lagrangian involving the Dirac
composite fermion field $\psi$, the emergent gauge field $a$, and the
external gauge field $A$ (the Chern-Simons term $-\frac1{4\pi}Ada$ is
included in $S_{\rm CF}$).  We assume for simplicity that $S_0$
is linear in $A_0$.  Since the electron density is
\begin{equation}
  \rho = \frac{\delta S_0}{\delta A_0} + \frac12
         \frac{\delta S_{\nu=1}}{\delta A_0} \,,
\end{equation}
the third term on the right-hand side of Eq.~(\ref{S-improved}) is the
interaction energy.

In Eq.~(\ref{S-improved}) we have included an infinite number of local
interaction terms between the composite fermion charge density $\delta
S_0/\delta A_0$ and the magnetic field.  Since these terms are all
local, nothing prevents them from arising in the low-energy effective
theory.  We now rewrite the action to the form
\begin{equation}
  S = S_0(\psi,a,\tilde A)
      - \frac V2\left(\frac{\delta S_0}{\delta A_0}\right)^2
      - \frac V8 \left(\frac{\delta S_{\nu=1}}{\delta A_0}-\rho_0\right)^2
      + \frac12 S_{\nu=1}[A],
\end{equation}
where $\rho_0$ is the background density and
\begin{equation}
  \tilde A_0 = A_0 - \frac V2 \left(\frac{\delta S_{\nu=1}}{\delta A_0}
  -\rho_0 \right)
     + \sum_{n=0}^\infty C_n \nabla^{2n} \delta B.
\end{equation}
To linear order,
\begin{equation}
  \tilde A_0(q) = A_0(q) - \left[
    \frac1{4\pi}\frac2{q^2}(1-e^{-q^2/2}) V(q) - \sum_{n=0}^\infty
    (-1)^n C_n q^{2n}
    \right]\delta B.
\end{equation}
Now if $C_n$ are related to the coefficients $V_n$ in
Eq.~(\ref{Fex-series2}) by
\begin{equation}
  C_n = \frac{V_{n+1}}{2^{n+1}(n+1)!}\,,
\end{equation}
then one recognizes the expression in the square brackets to be $\tilde V(q)$, so
\begin{equation}
  \tilde A_0 = A_0 - \frac{\tilde V\ast\delta B}{4\pi}\,.
\end{equation}
Again, the action can be broken into two parts,
\begin{equation}
   S = S_{\rm ph}[\psi, a, \tilde A] + \frac12 S_{\nu=1}[A],
\end{equation}
with the $S_{\rm ph}$ part having the symmetry~(\ref{mod_CT}).  From
this we find the exact relation~(\ref{exact-res}).  We conclude that
if the action can be brought to the form~(\ref{S-improved}), where
$S_0$ depends linearly on $A_0$ and has the symmetry~(\ref{CT}), then
the exact relationship between $\sigma_H$ and $\chi$ is guaranteed.

At this moment we do not know how to write the
action~(\ref{S-improved}) in a more compact and more natural form.  We
expect additional terms to appear if one goes to higher orders in
$\delta B$.  One should be able to write these terms down by repeating
the calculations of Sec.~\ref{sec:PH-nonuniform} beyond linear
order in $\delta B$.

\section{Conclusion and comments on the HLR theory}

We have shown that particle-hole symmetry of the lowest Landau level
leads to an exact relationship between the Hall conductivity and the
susceptibility.  The relationship is valid at all wave numbers and for
all frequencies (much smaller than the cyclotron frequency,
but the frequency can be in any relationship with the interaction energy)
in the absence of
disorder or in the presence of statistically particle-hole symmetric
disorder.  This relationship should be verifiable in numerical
simulations.  For the latter, the Hall conductivity can be found,
e.g., by measuring the density in inhomogeneous magnetic field.

We have shown that the exact relationship requires that the action of
the Dirac composite fermion theory contains an infinite number of
local terms, with coefficients fixed by the two-body electron-electron
potential.  This seems to be related to a discrete symmetry and thus
should not be viewed as fine-tuning.

The HLR field theory does not have an explicit particle-hole symmetry,
and hence \emph{a priori} it is not clear if it can be modified so that the exact
relationship~(\ref{exact-res}) holds for all $\omega$ and $q$.  While a
full analysis is still to be made, one problem can already be identified
in the limit of small
$\omega$ and $q$, $q/\omega\to0$, where 
Eq.~(\ref{sH-smallq}) predicts the value for the coefficient of the
$q^2$ correction to the Hall
conductivity.
One might think that this coefficient
can be tuned to any value
by adding higher-derivative terms to the HLR Lagrangian.  However,
this is not true.  It is known that Galilean invariance relates this
coefficient to the (frequency-dependent) Hall viscosity $\eta_H$ at
zero $q$~\cite{Hoyos:2011ez,Bradlyn:2012ea}.  For the $g=2$
current considered in this paper, the relationship reads
\begin{equation}
  \sigma_H(\omega,q) = \frac\nu{2\pi} + \left(\eta_H(\omega) -\frac\rho2\right)
    q^2 + o(q^2),
\label{sigmaetarel}
\end{equation}
where $\rho$ is the particle number density.  The Hall viscosity is
related to the average orbital spin per particle $s$
by~\cite{Read:2010epa}
\begin{equation}
  \eta_H = \frac{\rho s}2\,.
\label{etaspinrel}
\end{equation}
Equation~(\ref{sH-smallq}) then translates to an average orbital spin
$s=\frac12$ per particle in the Fermi-liquid state, exactly the same
value as in the filled Landau level.  On the other hand, in the HLR
theory the composite fermion acquires orbital spin from flux
attachment: each unit of flux increases the orbital spin of the
composite particle by $\frac12$~\cite{Cho:2014vfl}, making the
composite fermion of the half-filled Landau level have orbital spin
1, which differs from the value we have just deduced from particle-hole
symmetry and which would lead to a vanishing $q^2$ correction to $\sigma_H$.

One may wonder if it is possible to modify the HLR theory so that the
composite fermion would have orbital spin $\frac12$ instead of 1.
However, such a modification would break the orbital spin of the
gapped states derived from the HLR Fermi liquid state: the
Jain-sequence states $\nu=\frac n{2n+1}$ and $\nu=\frac{n+1}{2n+1}$,
and the Pfaffian state.  The $\nu=\frac n{2n+1}$ state is obtained from the
HLR state by placing the composite fermions into $n$ filled Landau
levels, increasing the orbital spin per fermion by $\frac n2$ to a
total of $1+\frac n2=\frac{n+2}2$, consistent with the shift~\cite{Wen:1992ej} $n+2$ for
this state.  In the PH conjugate state $\nu=\frac{n+1}{2n+1}$ the CFs
fills $n+1$ LLs in a magnetic field of opposite sign, making the
orbital spin $1-\frac{n+1}2=\frac{1-n}2$, again matching the shift
$1-n$.  In the Moore-Read state the composite fermions form spin-1
Cooper pairs in which each fermion receives an additional orbital spin
$\frac12$ for a total of $1+\frac12=\frac32$, matching the shift
$\mathcal{S}=3$.  Any modification of the HLR theory that changes the
orbital spin of the CF would destroy these agreements.

To further illustrate the difficulties of the HLR theory with particle-hole symmetry, in the Appendix we consider a concrete realization of this
theory: an approximation scheme developed in
Refs.~\cite{SimonHalperin:1993,SimonSternHalperin:1996} under the name
``magnetized modified random-phase approximation'' (MMRPA).  We show that this approximation
accurately reproduces the $q^2$ term in the dc Hall conductivity of
the Jain-sequence states, but fails to get the PH-symmetric
coefficient of this term in the ac Hall conductivity of the Fermi-liquid state.

\acknowledgments

The authors thank Andrey Gromov, Ilya Gruzberg, Dung Xuan Nguyen,
Matthew Roberts, and Chong Wang for discussions.  This work is
supported by the NSF under grant No.\ DMR-1254741, the DOE under grant No.\ DE-FG02-13ER41958, and the Simons Foundation.

\appendix

\section{An HLR mean-field calculation of the $q^2$ correction
  to Hall conductivity} \label{mmrpa_app}

We start from the HLR effective theory,
\begin{equation}\label{HLR}
  S = S[\psi, A-a] + \frac1{8\pi}ada
      + \frac V{2(4\pi)^2} (\nabla\times a)^2,
\end{equation}
where $S[\psi,A-a]$ is the action describing the coupling of the
nonrelativistic CF with the gauge field $(A-a)_\mu$.
Integrating over $\psi$
and keeping only the quadratic term, this action is replaced by $\frac12
(A-a)\cdot\Pi_{\rm CF}\cdot(A-a)$, where $\Pi_{\rm CF}^{\mu\nu}$ is the
electromagnetic response function of the CF.  Working in the temporal
gauge, this matrix becomes a $2\times2$ matrix of the spatial
components.  In this gauge it is easy to integrate out $a$ to get the
response matrix of the electron $\Pi$,
\begin{equation}\label{RPA}
  \Pi = - Q
  \left(\Pi_{\rm CF}+ Q \right)^{-1} Q
    + Q,
\end{equation}
where the matrix $Q$ is defined as
\begin{equation}
  Q_{ij} = \frac{i\omega}{4\pi}\epsilon_{ij}- \frac{V(q)}{(4\pi)^2}
  (q^2\delta_{ij}-q_iq_j).
\end{equation}

In the most naive random-phase approximation (RPA), $\Pi_{\rm CF}$ is assumed to be the response
function of a free composite Fermi gas $\Pi_{\rm CF}^{\rm free}$
computed assuming the CF has effective mass $m_*$.  This approximation
breaks Galilean invariance when $m_*$ differs from the bare electron
mass $m$ and hence cannot be used if one wants to take the LLL
$m\to0$.  To correct the problem, a modification of the RPA was
proposed in Ref.~\cite{SimonHalperin:1993} which was later amended in
Ref.~\cite{SimonSternHalperin:1996} into a scheme called MMRPA.
We reinterpret this modification as replacing $S[\psi,A-a]$ not by an
action of a free fermion coupled to $a_\mu$ but by
\begin{equation}\label{S-MRPA}
  S[\psi, A-a, v^i] = i\psi^\+ D_t\psi - \frac{|D_i\psi|^2}{2m(1+F_1)}
  + \frac{F_1}{1+F_1}\frac i2 v^i \psi^\+\Dlr_i\psi
  + \frac{F_1}{1+F_1}\frac{mv^2}2 \psi^\+\psi,
\end{equation}
where $v^i$ is a field to be integrated over and $F_1$ is a
parameter.  This action describes a theory of a fermion with an
effective mass $m_*=m(1+F_1)$ but is consistent with the Galilean
invariance of the original electrons with mass $m$.  One quick way to
see that is to follow Ref.~\cite{Greiter:1989qb} to check, with the
help of the equation of motion $\delta S/\delta v^i=0$, that the
momentum density $T^{0i}$ is the particle number $j^i$ times the bare
electron mass $m$:
\begin{equation}
  \frac{\d\mathcal{L}}{\d(\d_t\psi)} D_i\psi
  +  D_i\psi^\+ \frac{\d\mathcal{L}}{\d(\d_t\psi^\+)}
  = m\frac{\d\mathcal{L}}{\d A_i} \,.
\end{equation}
If one integrate out $v^i$ first, one generates a contact interaction
for the fermion.  Hence one can interpret $F_1$ as
the $p$-wave Landau parameter.  The LLL limit corresponds to taking
$F_1\to\infty$, with $m_*=m(1+F_1)$ fixed.  In this limit the last term on
the right-hand side of Eq.~(\ref{S-MRPA}) disappears.

The scheme developed in
Refs.~\cite{SimonHalperin:1993,SimonSternHalperin:1996} is essentially
the RPA in the theory~(\ref{HLR}) and (\ref{S-MRPA}).  One integrates out
$\psi$ first and keep only terms quadratic in $a$ and $v$,
\begin{equation}
  \frac12 [A-a-(m_*-m)v]\cdot \Pi_{\rm CF}^{\rm free}\cdot [A-a-(m_*-m)v]
  + \frac\rho 2 (m_*-m)v^2,
\end{equation}
where $\rho$ is the particle number density.
Then one performs the Gaussian integration over $v^i$ in the Gaussian
approximation to obtain $\frac12 (A-a)\cdot \Pi_{\rm CF}\cdot (A-a)$, where
\begin{equation}\label{MRPA}
  \Pi_{\rm CF} = \left(1 + \frac{m_*-m}\rho\Pi_{\rm CF}^{\rm free}\right)^{-1}
      \Pi_{\rm CF}^{\rm free}.
\end{equation}
This is exactly the prescription of the modified
RPA~\cite{SimonHalperin:1993}.  The later improvement, magnetized
modified RPA~\cite{SimonSternHalperin:1996} simply declares the
$\Pi$ obtained in Eq.~(\ref{RPA}) to corresponds to the $g=2$
electromagnetic response.

From now on we choose $\q$ to point along the $x$ axis, $\q=(q,0)$.
For the $\nu=\frac n{2n+1}$ state, the composite fermions live in a
reduced magnetic field $b=\frac B{2n+1}$.  In the small-$\omega$,
small-$q$ limit, $\omega/q\to0$, the response function of a free gas
is~\cite{Chen:1989xs}
\begin{align}
  \left(\Pi^{\rm free}_{\rm CF}\right)_{11} &= \frac n{2\pi} \frac{m_*}b \omega^2 + O(\omega^2q^2),\\
  \left(\Pi^{\rm free}_{\rm CF}\right)_{12} &= i\omega \frac n{2\pi} \left(1-\frac{3n}4 (q\ell_b)^2\right) + O(\omega q^4),\\
  \left(\Pi^{\rm free}_{\rm CF}\right)_{21} &=
  -\left(\Pi^{\rm free}_{\rm CF}\right)_{12},\\
  \left(\Pi^{\rm free}_{\rm CF}\right)_{22} & = -\frac{n^2}{2\pi} \frac{q^2}{m_*} + O(q^4),
\end{align}
where $\ell_b^2=1/b=(2n+1)\ell_B^2$.  Inserting these formulas into
Eqs.~(\ref{RPA}) and (\ref{MRPA}), one finds the $q$ dependence of the dc Hall conductivity $\sigma_H=-i\lim_{\omega_\to0}\omega^{-1}\Pi_{12}$ to be
\begin{equation}
  \sigma_H(q) = \frac1{2\pi} \frac n{2n+1} \left(1+ \frac n4 q^2\ell_B^2\right).
\end{equation}
This agrees with the general formula
\begin{equation}
  \sigma_H(q) = \frac\nu{2\pi} \left(1+ \frac{\mathcal{S}-2}4 q^2\ell_B^2\right)
\end{equation}
when one substitutes in the latter the value of the shift of the Jain
state, $\mathcal{S}=n+2$.

Analogously, for the $\nu=\frac{n+1}{2n+1}$ state,
\begin{align}
  \left(\Pi^{\rm free}_{\rm CF}\right)_{11} &= \frac {n+1}{2\pi} \frac{m_*}{|b|} \omega^2 + O(\omega^2q^2),\\
  \left(\Pi^{\rm free}_{\rm CF}\right)_{12} &= -i\omega \frac{n+1}{2\pi} \left(1-\frac{3(n+1)}4 (q\ell_b)^2\right) + O(\omega q^4),\\
  \left(\Pi^{\rm free}_{\rm CF}\right)_{21} &=
  - \left(\Pi^{\rm free}_{\rm CF}\right)_{12},\\
  \left(\Pi^{\rm free}_{\rm CF}\right)_{22} & = -\frac{(n+1)^2}{2\pi} \frac{q^2}{m_*} + O(q^4),
\end{align}
and after some calculation one obtains
\begin{equation}
  \sigma_H = \frac1{2\pi} \frac{n+1}{2n+1} \left( 1-
  \frac{n+1}4 q^2\ell_B^2\right),
\end{equation}
which matches the value of the shift $\mathcal{S}=-n+1$.

Let us now turn to the Fermi-liquid state.  For small $\omega$ and
$q$ but $\omega/v_Fq\gg1$, the response function of a Fermi gas is
\begin{align}
  \left(\Pi^{\rm free}_{\rm CF}\right)_{11} &= -\frac\rho{m_*} \left( 1
  + \frac34\frac{v_F^2q^2}{\omega^2}\right),\\
  \left(\Pi^{\rm free}_{\rm CF}\right)_{12} &= 
  \left(\Pi^{\rm free}_{\rm CF}\right)_{21} = 0, \\
  \left(\Pi^{\rm free}_{\rm CF}\right)_{22} &= - \frac\rho{m_*} \left( 1
  + \frac14\frac{v_F^2q^2}{\omega^2}\right).
\end{align}
Applying formulas~(\ref{RPA}) and (\ref{MRPA}) we find
\begin{equation}
  \sigma_H = \frac1{4\pi} + O(q^3),
\end{equation}
with vanishing coefficient in front of the $q^2$ term,
which does not match the requirement of particle-hole symmetry~(\ref{sH-smallq}).

One can also put the action~(\ref{HLR}) and (\ref{S-MRPA}) in curved
space by using the metric tensor $g^{ij}$ to sum over spatial indices
and replacing the covariant derivative by
$D_\mu\psi=(\d_\mu-iA_\mu+ia_\mu-is\omega_\mu)$, with $s$ being the orbital
spin of the composite fermion and the spin connection $\omega_\mu$
defined as in Ref.~\cite{Hoyos:2011ez} so that it vanishes in flat
space and any electromagnetic field.  It happens that for $s=1$ one
does not need to introduce any higher-order term to make the theory
consistent with the nonrelativistic general coordinate invariance of
the original electron theory with $g=2$~\cite{Geracie:2014nka}.  This
explains why the MMRPA reproduces correctly the $O(q^2)$ correction to
the Hall conductivity for the Jain sequence at $\nu = \frac{n}{2n \pm 1}$.  It also implies that for Jain states around
other even-denominator filling fractions, e.g., $\nu=\frac14$, the
unmodified MMRPA will not give the correct $q^2$ correction to
$\sigma_H$.

\end{document}